\documentclass[twocolumn,showpacs,preprintnumbers,amsmath,amssymb]{revtex4}
\usepackage{graphicx}% Include figure files
\usepackage{dcolumn}% Align table columns on decimal point
\usepackage{bm}% bold math
\begin{document}
\widetext

\title{Dependence of the superconducting transition temperature on the doping level in single crystalline diamond films }

\author{E. Bustarret$^1$, J. Ka\v{c}mar\v{c}ik$^{2,3}$, C. Marcenat$^3$ , E.
Gheeraert$^1$, C. Cytermann$^4$, J. Marcus$^1$ and T. Klein$^{1,5}$}
\address{$^{1}$ Laboratoire d'Etudes des Propri\'et\'es Electroniques
des
Solides, CNRS, B.P.166, 38042 Grenoble Cedex 9, France}
\address{$^2$  Centre of  Low Temperature  Physics IEP  SAS \& FS
UPJ\v S, Watsonova 47, 043 53 Ko\v{s}ice, Slovakia   }
\address{$^{3}$ Commissariat \`a l'Energie Atomique - Grenoble,
D\'epartement de Recherche Fondamentale sur la Mati\`ere Condens\'ee,
SPSMS, 17 rue des Martyrs, 38054 Grenoble Cedex 9, France}
\address{$^{4}$ Physics Department and Solid State Institute, Technion, 32000 Haifa, Israel}
\address{$^{5}$ Institut Universitaire de France and Universit\'e Joseph
Fourier, B.P.53, 38041 Grenoble Cedex 9, France}
\date{\today}

\begin{abstract}
Homoepitaxial diamond layers doped with boron in the
$10^{20}-10^{21}$ cm$^{-3}$ range are shown to be type II
superconductors with sharp transitions ($\sim 0.2$ K) at  temperatures increasing from $0$ to $2.1$ K with boron contents. The critical concentration for the onset of
superconductivity is about $5-7$ $10^{20}$ cm$^{-3}$, close to the metal-insulator transition. The $H-T$ phase diagram has been obtained from transport and a.c. susceptibility measurements down to $300$ mK.  These results bring new quantitative constraints on the theoretical models proposed for superconductivity in diamond.
\end{abstract}

\maketitle

Type II superconductivity has been recently reported for heavily
boron-doped polycrystalline diamond prepared either as bulk
\cite{Ekimov} or thin film samples  \cite{Takano}, providing a new
interesting system for the study of superconductivity in doped
semiconductors. Based on the great bonding strength of the valence
band states and on their strong coupling to the carbon lattice phonons,
various theoretical studies \cite{Boeri,Lee} have stressed the
similarity between diamond and the recently discovered MgB$_2$ system
which shows a surprizingly high $T_c$ value on the order of $40$ K.
Those calculations lead to $T_c$ values in the $0.2$ to $25$ K range (depending on
the boron content) but ignore the boron impurity band
\cite{Boeri,Lee}. An alternative theoretical approach \cite{Baskaran}
stresses out the fact that the boron concentration range where
superconductivity has been observed is close to the
Anderson-Mott metal-insulator transition and
suggests an electron correlation driven extended $s-$wave
superconductivity in the boron impurity band.\

The open questions of the nature of the metal-insulator transition (MIT) in diamond and of its correlation with superconductivity are of fundamental interest and provide ample motivation for this first investigation of the dependence of the superconducting transition temperature on the doping boron concentration. In this letter, we report on magnetic and transport experiments on a set of high quality single crystalline epilayers doped in the
relevant $10^{20}-10^{21}$ cm$^{-3}$ range. We show that $T_c$
rapidly increases above some critical concentration $\sim 5-7$
$10^{20}$ cm$^{-3}$ reaching $\sim 2$ K for $n_B = 19$ $10^{20}$
cm$^{-3}$ (see Table 1).

\begin{table}
\caption{\label{table1} Sample characteristics : thickness ($t$), gas phase ratio ((B/C)$_{gas}$), boron concentration ($n_B$) and critical temperatures ($T_c$) for the studied diamond epitaxial films }
\begin{ruledtabular}
\begin{tabular}{ccccc}
Sample&$t$ ($\mu$m)& (B/C)$_{gas}$ (ppm)&$n_B$ ($10^{20}$ cm$^{-3}$)&$T_c$ (K)\\
\hline
1 & 3.0 & 1615 & 3.6 & $\leq$ 0.05\\
2 & 3.0 & 1730 & 9 & 0.9\\
3 & 3.0 & 1845 & 10 & 1.2 \\
4 & 2.0 & 2200 & 11.5 & 1.4 \\
5 & 0.15 & 2800 & 19 & 2.1 \\
\end{tabular}
\end{ruledtabular}
\end{table}

\begin{figure}
\begin{center}
\resizebox{0.5\textwidth}{!}{\hskip -1cm\includegraphics[angle=270]{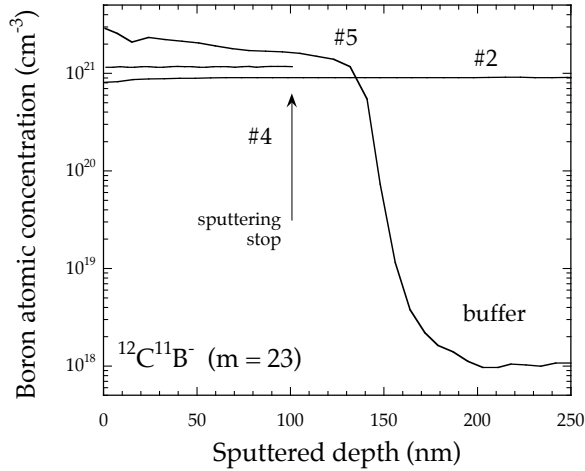}}
\caption{SIMS profiles for ion mass $23$ obtained using $Cs^+$ primary ions on samples 2, 4 and 5. In the case of the thicker samples, sputtering was interrupted before reaching the buffer layer.}
\label{Fig.1}
\end{center}
\end{figure}

{001}-oriented type Ib diamond substrates were first exposed to a pure
hydrogen plasma. Methane (4\%) was subsequently introduced and a $0.5
\mu$m-thick buffer layer of non-intentionally doped material was
deposited at $820^\circ$ C by the microwave plasma-assisted decomposition (MPCVD) of
the gas mixture at a total pressure of $30$ Torr. Finally, diborane was
introduced in the vertical silica wall reactor with boron to carbon
concentration ratios in the gas phase ($(B/C)_{gas}$)ranging from $1500$ to $3000$
ppm. With a typical growth rate of $0.9 \mu$m/hr these deposition
conditions led to $0.1$ to $4 \mu$m-thick $p^+-$type diamond layers. 
Secondary Ion Mass Spectroscopy (SIMS) depth profiles of $^{11}B^-$, $^{12}C^-$ and $^{11}B^{12}C^-$ ions were measured using a $Cs^+$ primary ion beam in a Cameca Ims $4f$ apparatus. As shown in Fig.1 for samples 2, 4 and 5, these profiles were found to be flat or with a slow decrease toward the buffer layer. The boron atomic densities $n_B$ shown both in Fig.1 and Table 1 were derived from a quantitative comparison to a SIMS profile measured under the same conditions in a $B-$implanted diamond crystal with a known peak boron concentration of $2.4$ $10^{20}$ cm$^{-3}$. For thin enough samples, the profile yielded also the residual doping level in the buffer layer, around $10^{18}$ cm$^{-3}$. Moreover, the single crystal and epitaxial character of the MPCVD layers was checked by high resolution X-ray diffraction, yielding shifted narrow lines with smaller linewidths at half maximum than for the Ib substrate (typically $10$ arcsec for the {004} Bragg spot of the epilayer, instead of $13$ arcsec for the substrate).
The chemical composition \cite{Bustarret1}, structural
\cite{Bustarret2,Brunet} and optical \cite{Bustarret1,Brunet,Pruvost}
characteristics of  these layers at room temperature are well
documented and have been reviewed recently, together with some
preliminary transport measurements \cite{Brunet}.

The superconducting temperatures have been deduced from
ac-susceptibility measurements ($\chi_{ac}$). The films have been placed on top
of miniature coils and $T_c$ has been obtained by detecting the change
in the self induction of the coils induced by the superconducting
transitions. For fully screening samples, we observed a $4 \%$ drop of the
induction $L$ ($\sim 1$ mH). Small ac-excitation fields ($\omega \sim 99$ kHz and
$h_{ac} \sim $ a few mG) have been applied perpendicularly to the
films.  Four Au/Ti electrodes were deposited on top of the film with the
highest $T_c$ for magneto-transport measurements. A very small current
($\sim 1 nA$) corresponding to a current density on the order of
$10^{-3}$ Acm$^{-2}$ has been used to avoid flux flow dissipation. A standard lock-in technique at $17$ Hz was used to measure the temperature dependencies of the sample resistance at fixed magnetic fields. The measurements were performed down to $50$ mK ($\chi_{ac}$) and $300$mK (transport) and the magnetic field was applied perpendicularly to the doped plane of the sample, 

The critical temperatures (see Table 1) reported in Fig.2 have been
deduced from the onset of the diamagnetic signal (see inset of Fig.2).
The susceptibility has been rescaled to $-1$ in the superconducting
state and a $N \sim 0.9$ demagnetization coefficient has been used (the
values of $T_c$ do not depend on this choice). This choice for $N$
leads to a dissipation peak on the order of $0.2-0.3$ as expected for
the non linear regime. This is consistent with the observation of a
slight increase of the width of the transition for increasing $h_{ac}$
values. As shown in the inset of Fig.2 (and Fig.3), the transitions are very sharp
(with a width $\sim 0.2$ K) allowing an accurate determination of $T_c$
and stressing out the high quality of our samples. In comparison, the polycrystalline samples measured in previous reports presented a much larger resistivity transition width,  $\sim 1.7$ K in
\cite{Ekimov} and $\sim 2.6$ K in \cite{Takano}.

\begin{figure}
\begin{center}
\resizebox{0.5\textwidth}{!}{\hskip -1cm\includegraphics[angle=270]{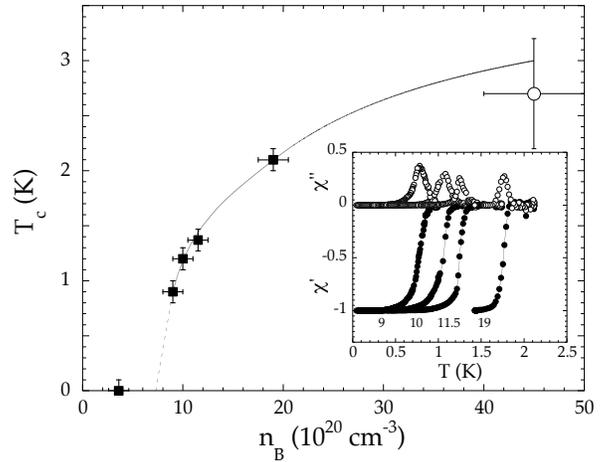}}
\caption{Dependence of the superconducting transition temperature $T_c$ on the boron concentration $n_B$ : closed squares : this work, open circle : from [1].The different $T_c$ values were obtained from the onset of the diamagnetic signal (see inset for the real (closed symbols) and imaginary (open symbols) parts of the magnetic susceptibility.}
\label{Fig.2}
\end{center}
\end{figure}

No transition was observed down to $50$ mK for the film
with $n_B = 3.6$ $10^{20}$ cm$^{-3}$. For higher boron
concentrations, $T_c$ increases rapidly with doping above some critical
concentration $\sim 5-7$ $10^{20}$ cm$^{-3}$ reaching $2.1$ K
for $n_B = 19$ $10^{20}$ at.cm$^{-3}$. The dependance of $T_c$ with
doping extrapolates towards the data recently obtained by
\cite{Ekimov}. On the contrary, our $T_c$ value for $n_B \sim 10^{21}$
cm$^{-3}$ ($\sim 1$ K) is much lower than the one recently reported
by \cite{Takano} ($\sim 4-7$ K). Our data suggests that these authors
have largely underestimated the boron concentration of their polycrystalline samples. They deduced this concentration from Hall measurements which are known to give results that deviate significantly from the actual boron concentration in $p^+-$type diamond \cite{Bustarret2}. It is also worth noticing that we observed superconducting transitions with $T_c$ on the order of $1$ K for boron contents $\sim
0.5 at.\%$ whereas recent calculations of the electron-phonon coupling led to much
smaller $T_c$ values for these doping levels  \cite{Boeri,Lee,Baskaran}.

\begin{figure}
\begin{center}
\resizebox{0.55\textwidth}{!}{\hskip -1cm\includegraphics[angle=0]{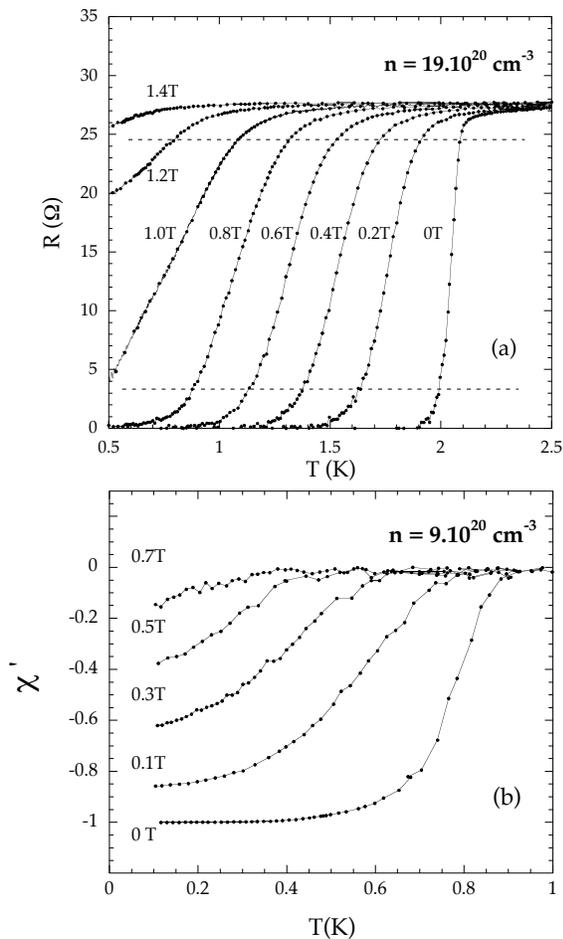}}
\vskip -1cm
\caption{(a) : temperateure dependence of the electrical resistance at indicated magnetic fields for the film with $n = 19.10^{20}$ at.cm$^{-3}$. The dashed lines correspond to the two criteria used for the determination of $H_{c2}$ (see Fig.3) i.e. $R/R_N = 90\%$ and $10\%$ ($R_N$ being the normal state resistance).(b) temperature dependence of the real part of the magnetic susceptibility at indicated magnetic fields for the film with $n =9.10^{20}$ at.cm$^{-3}$. $H_{c2}$ has been deduced from the onset of the diamagnetic signal.}
\label{Fig.3}
\end{center}
\end{figure}

The influence of an external field on the superconducting transition is
displayed in Fig.3. As shown in  panel a, the transition is shifted
towards lower temperatures as the magnetic field is increased. The
transition width remains relatively small up to $\sim 1$ T and rapidly
increases for larger fields. In the absence of thermodynamic
measurements, some care should be taken in order to define an accurate $H_{c2}$ line. This
line has been defined from the classical $R/R_n = 90 \%$ criterion
(where $R_N$ is the normal state resistance). As shown in Fig.4, the corresponding $H_{c2}(T)$ line can be well described by the classical
WHH theory \cite{WHH}. We hence get $H_{c2}(0) \sim 1.4$ T
corresponding to a coherence length $\xi_0 = \sqrt{\Phi_0/2\pi H_{c2}(0)} \sim 150 \AA$ for $n_B = 19$
$10^{20}$ cm$^{-3}$ ($\Phi_0$ being the flux quantum). We have also reported on Fig.4 the line corresponding to $R/R_n = 10 \%$ which gives an indication for the width of the transition, pointing out that the transition curves rapidly increase for $H > 1$ T. For sample 2, this line has been deduced from the shift of the diamagnetic response with increasing
fields. In this case, the rapid broadening of the transition is the hallmark of a
small critical current density ($J_c$) again emphasizing the high quality of our films. Indeed, in the non linear regime, the susceptibility is directly related to $J_cd/h_{ac}$ (where
$d$ is a characteristic length scale on the order of the sample
thickness) and $J_c \approx h_{ac}/d \sim 1$ Acm$^{-2}$ for $\chi '
\sim -0.4$. In this case, no saturation was observed down to
$200$ mK, indicating some deviation from the classical WHH behaviour.
Note that for both samples  the slope $[dH_{c2}/dT]_{T\rightarrow0}
\sim 1$ T/K, which is almost $2$ times smaller than the value reported
by Ekimov {\it et al.} \cite{Ekimov}.

\begin{figure}
\begin{center}
\resizebox{0.55\textwidth}{!}{\hskip -1cm\includegraphics[angle=270]{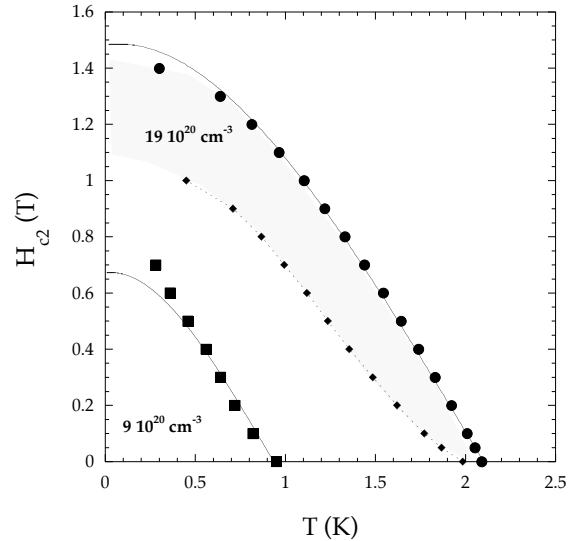}}
\caption{$H-T$ phase diagram for the films with $n_B = 19$ $10^{20}$ cm$^{-3}$ and $n_B = 9$ $10^{20}$ cm$^{-3}$. The circles (resp. diamonds) have being deduced from temperature sweeps of the electrical resistance (see fig.2) for $R/R_n = 90\%$ (resp. $R/R_N = 10\%$). The shadowed area is an indication of the width of the transition. The closed squares were defined from the onset of the ac-susceptibility (see Fig.2(b)). The full lines are fits to the data using the classical WHH theory.  }
\label{Fig.4}
\end{center}
\end{figure}

We obtained an almost temperature-independent normal state resistivity
on the order of $0.5$ m$\Omega$.cm for $n_B = 19$ $10^{20}$ cm$^{-3}$  suggesting that
our samples are close to the metal-insulator transition. On the basis of the criterion first proposed by Mott \cite{Mott} for metal-non metal transitions, which in its final form ($N_c^{1/3}a_H = 0.26$, where $a_H$ is the Bohr radius) has been verified in a wide variety of condensed media \cite{Edwards}, the critical concentration in $p-$type diamond is expected to be around $N_c = 1-2$ $10^{20}$ cm$^{-3}$ \cite{Shiomi}. Limitations to this approach arise from the discrepancies in the values  proposed in the literature for the acceptor Bohr radius $a_H$, in line with other inconsistencies about the valence band parameters of diamond. From an experimental point of view, extrapolations \cite{Borst} of the Pearson-Bardeen model \cite{Pearson} have led to $N_c$ values ranging from $1.5$ to $3.0$ $10^{20}$ cm$^{-3}$, while different transport measurements have prompted other authors to propose that $N_c \sim 7$ $10^{20}$ cm$^{-3}$ \cite{Nishimura}. It is worth noticing that those values have been deduced from room temperature measurements. To the best of our knowledge, the only low temperature estimate ($N_c \sim 40$ $10^{20}$ cm$^{-3}$) has been proposed by Tschepe {\it et al.} \cite{Tschepe}  on the basis of a scaling analysis of the electrical conductivity. However, the absolute conductivity values of their implanted samples are much lower than ours despite larger boron concentrations.\ 

The proximity of a MIT in this system, makes it very favorable for the
observation of quantum fluctuation effects. The strength of these
fluctuations can be quantified through the quantity $Q_u = R_{eff}/R_Q$ where $R_Q=\hbar/e^2  \sim 
4.1$ k$\Omega$ is the quantum resistance and $R_{eff}=\rho_N/s$ ($s$ is a relevant length scale for these fluctuations  \cite{Blatter}).  Taking $\rho_N \sim 5$ $10^{-4} \Omega$.cm and $s \sim \xi (0) \sim 150 \AA$,  we obtain a large $Q_u$ ratio $\sim 0.1$ indicating
that quantum effects  may be important in this system. These quantum fluctuations may give rise to the melting of the flux line lattice and
thus lead to the rapid broadening of the resitive transitions observed at low
$T$ and large $H$. Another indication for such quantum effects, is the
almost temperature-independent mixed state resistivity above $1.2$ T as
previously observed in other systems with similar $Q_u$values \cite{Yazdani,Okuma}

To conclude, we were able to prepare highly homogeneous and well characterized boron-doped diamond films in the $10^{20}$ - $10^{21}$ cm$^{-3}$ range where superconductivity occurs. The value of the critical concentration for the onset of superconductivity is on the order of $5-7$ $10^{20}$ cm$^{-3}$. Boron-doped diamond is an ideal system to study the occurence of superconductivity close to the metal-insulator transition. As a consequence, quantum effects are expected to play a significant role as suggested by the large quantum resistance ratio $Q_u \sim 0.1$.

We are grateful to F. Pruvost who grew some of theÊ diamond layers and to Dr L. Ortega (Cristallographie, CNRS, Grenoble) for the X-ray diffraction experiments.

\end{document}